\documentclass[aps,showpacs,twocolumn,superscriptaddress,showpacs,amsmath,amssymb,prl]{revtex4} %
\usepackage{graphicx,graphics}
\def\gsim{\mathrel{\rlap{\lower4pt\hbox{\hskip1pt$\sim$}}
    \raise1pt\hbox{$>$}}}
\def\lsim{\mathrel{\rlap{\lower4pt\hbox{\hskip1pt$\sim$}}
    \raise1pt\hbox{$<$}}}

\begin{document}
\title{The ratchet effect and
the transporting islands in the chaotic sea}

\author{Lei Wang}
\affiliation{Department of Physics and Centre for Computational
Science and Engineering, National University of Singapore,
Singapore 117542, Republic of Singapore}

\author{Giuliano Benenti}
\affiliation{Center for Nonlinear and Complex Systems, Universit\`a degli
Studi dell'Insubria, Via Valleggio 11, 22100 Como, Italy}
\affiliation{CNISM, CNR-INFM, and Istituto Nazionale di Fisica
Nucleare, Sezione di Milano}

\author{Giulio Casati}
\affiliation{Center for Nonlinear and Complex Systems, Universit\`a degli
Studi dell'Insubria, Via Valleggio 11, 22100 Como, Italy}
\affiliation{CNISM, CNR-INFM, and Istituto Nazionale di Fisica
Nucleare, Sezione di Milano}
\affiliation{Department of Physics and Centre for Computational
Science and Engineering, National University of Singapore,
Singapore 117542, Republic of Singapore}

\author{Baowen Li}
\affiliation{Department of Physics and Centre for Computational
Science and Engineering, National University of Singapore, Singapore
117542, Republic of Singapore}
\affiliation{ NUS Graduate School of
Integrative Sciences and Engineering, 117597,
 Republic of Singapore}

\date{\today}
\begin{abstract}
We study directed transport in a classical deterministic dissipative
system. We consider the generic case of mixed phase space and show
that large ratchet currents can be generated thanks to the presence,
in the Hamiltonian limit, of transporting stability islands embedded
in the chaotic sea. Due to the simultaneous presence of chaos and
dissipation the stationary value of the current is independent of
initial conditions, except for initial states with very small measure.
\end{abstract}


\maketitle

The ratchet effect, that is the possibility of obtaining directed
transport of particles in the absence of any net bias force, is a
problem at the heart of statistical mechanics. Ratchet transport in
systems at equilibrium is forbidden by the second principle of
thermodynamics \cite{feynman}. On the other hand, it is possible to
overcome this limitation in systems out of equilibrium, provided all
space-time symmetries which inhibit directed motion are broken
\cite{flach}. The ratchet phenomenon has recently gained renewed
interest \cite{hanggi,reimann} as a model elucidating the physics of
molecular motors \cite{julicher}. Moreover, directed transport may
lead to technological applications at the nanoscale \cite{linke},
including new electron pumps, molecular switches, rectifiers and
transistors.

Previous theoretical studies have shown the ratchet effect in
systems in which noise is absent and its role is played by the
deterministic chaos induced by the inertial term \cite{jung}. In
particular the origin of current reversal in such inertia ratchets
has been carefully investigated \cite{mateos}.

In spite of these pioneering works, the interrelation between the
complexity and the rich variety of classical chaotic motion in
conservative systems and the appearance of the ratchet phenomenon
when dissipation is introduced, is not known. In particular the role
of stable islands in the mixed phase space structure which is {\it
generic} in nonlinear dynamical systems is not clear \cite{Hlimit}.

In this paper, by considering a periodically kicked, dissipative,
inertia ratchet, we show that the generic mixed phase space
structure of the conservative case may lead to a strong ratchet
phenomenon when dissipation is introduced. In particular, for strong
dissipation, large currents may arise in a short time scale. On the
other hand, for weak dissipation, large ratchet currents can be
achieved nearly independently of initial conditions, as a result of a
beautiful interplay between chaotic diffusion, ballistic transport
of islands and dissipation.

The system we study is a particle moving in one dimension
[$x$$\in$$(-\infty,\infty)$] in a periodic kicked asymmetric
potential:
\begin{align}
\begin{array}{l}
V(x,\tau)=K[\mbox{cos}(x)+\displaystyle\frac a2 \mbox{cos}(2x+\phi)]\sum_{m=-\infty}^{\infty} \delta(\tau-mT)
\end{array},
\label{ratchetpotential}
\end{align}
where $T$ is the kicking period, which is set to unity in this
paper. The evolution of the system in one period is described by the map
\begin{align}
\left\{ \begin{array}{l}
  \bar p=\gamma p+K[\mbox{sin}(x)+a \mbox{sin}(2x+\phi)],  \\
         \bar x=x+\bar p,
\end{array}\right.
\label{ratchetmap}
\end{align}
where $p$ is the momentum variable conjugated to $x$ and
$\gamma$$\in$$[0,1]$ is the dissipation parameter, describing a
velocity proportional damping. The limiting cases $\gamma=0$ and
$\gamma=1$ correspond to overdamping and Hamiltonian evolution,
respectively.  The spatial symmetry is broken at $a\ne 0$, $\phi\ne
n\pi$, with $n$ integer \cite{zaslavsky}. However directed transport
is forbidden in the Hamiltonian limit due to time-reversal symmetry
\cite{qratchet}. In this limit system (\ref{ratchetmap}) exhibits
the typical mixed phase space structure. A central point of this
paper is to show that even very tiny, almost invisible islands,
embedded in the chaotic sea play a crucial role in the generation of
large ratchet currents when a small dissipation is added.

\begin{figure}[ht]
\includegraphics[width=\columnwidth]{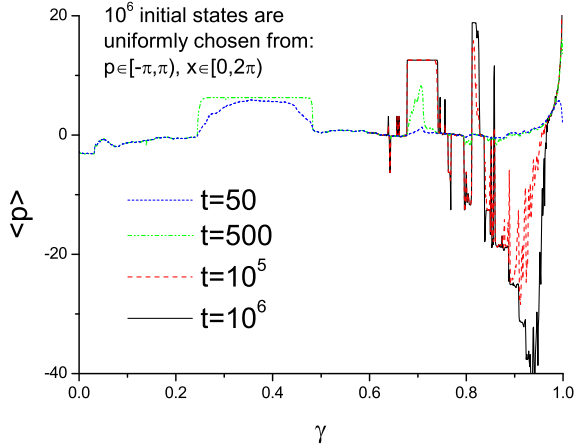}
\vspace{-1cm} \caption{\label{fig:ensembleaveragep_k65a05}
Ensemble average current $\langle p \rangle$ at different times,
for $a=0.5$.}
\end{figure}

In Fig.~\ref{fig:ensembleaveragep_k65a05} we plot the ratchet
current $\langle p \rangle$ as a function of the dissipation
parameter $\gamma$ at different times $t$ (the discrete time $t$
measures the number of kicks). Here we set $K=6.5$, $a=0.5$,
$\phi=\pi/2$ and we follow the evolution in time of a large number
of trajectories whose initial conditions are randomly and uniformly
distributed in the \emph{unit cell} $-\pi\le p <\pi$, $0\le x <
2\pi$. Therefore, the initial average momentum is $\langle p
\rangle=0$ for any $\gamma$. Since all relevant space-time
symmetries \cite{flach} are broken at $\gamma\ne 1$, then a directed
current $\langle p \rangle\ne 0$ can be generated. The dependence of
the current on $\gamma$ is rather complicated. In particular, we can
see plateau regions inside which the asymptotic ratchet current is
independent of $\gamma$. We note that for strong dissipation
($\gamma\lsim 0.6$) the current converges to its asymptotic value
very rapidly and typically independently of initial conditions, the
dynamics being characterized by a single stationary distribution.
The weakly dissipative regime will be discussed later.

In order to understand the behavior of the ratchet current we first
perform a linear stability analysis of map (\ref{ratchetmap}).
The fixed points of the map when $x$ is taken modulo $2\pi$
are given by
\begin{align}
\left\{ \begin{array}{l}
  p^{*}=2l\pi,      \mbox{ $l$ integer},  \\
  (\gamma-1)2l\pi+K[\sin(x^{*})+a \cos(2x^{*})]=0,
\end{array}\right.
\label{fixedpoints}
\end{align}
where for the sake of simplicity we have considered the case
$\phi=\pi/2$. The stability of these fixed points is determined by
the  eigenvalues $\lambda_{1,2}$ of the Jacobian
\begin{align}
  \frac {\partial(\bar p,\bar x)}{\partial(p,x)}=
\left( \begin{array}{cc}
  \gamma & K[\mbox{cos}(x)-2a \mbox{sin}(2x)]  \\
  \gamma & 1+K[\mbox{cos}(x)-2a \mbox{sin}(2x)]
\end{array} \right).
\label{stabmat}
\end{align}

\begin{figure}[ht]
\includegraphics[width=\columnwidth]{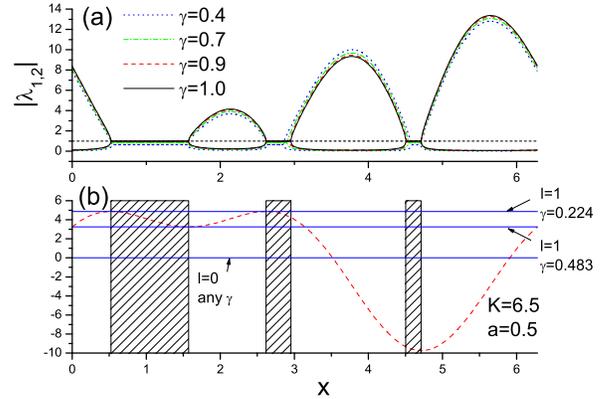}
\vspace{-1cm} \caption{\label{fig:fixed_point_k65a05} (a) Modules of
the eigenvalues $\lambda_{1,2}(x)$ of the stability matrix
(\ref{stabmat}) for different values of $\gamma$. (b) Dashed curve:
$f(x)$$=$$K[\mbox{sin}(x)$$+$$a \mbox{cos}(2x)]$; solid lines:
$g(x)$$=$$(1$$-$$\gamma)2l\pi$ for different $l$ and $\gamma$. In
the shadowed regions, $|\lambda_{1,2}(x)|$ are both less than one.}
\end{figure}

The modules $|\lambda_{1,2}|$ versus $x$ are shown in
Fig.~\ref{fig:fixed_point_k65a05}(a) for different $\gamma$. The
stable intervals, in which the modules of the eigenvalues are both
less than 1, depend only very slightly on $\gamma$. Therefore we may
consider the case $\gamma=1$, for which the stability intervals can
be computed analytically and are given by
$(\arcsin\displaystyle\frac1{4a},\displaystyle\frac{\pi}2)$,
$(\pi-\arcsin\displaystyle\frac1{4a},c_1)$,
$(c_2,\displaystyle\frac{3\pi}2)$ where $c_1,c_2$ are the real roots
of the equation $\mbox{cos}(x)-2a
\mbox{sin}(2x)=-\displaystyle\frac{4}{K}$.
These intervals are shown as shadowed regions in
Fig.~\ref{fig:fixed_point_k65a05}(b).

As illustrated in Fig.~\ref{fig:fixed_point_k65a05}(b), the fixed
points $({x^{*}},{p^{*}}=2\pi l)$ can be determined graphically.
If ${x^{*}}$
resides in a shadowed region, then the fixed point is stable.
Clearly there are no stable fixed points for $l$=0. Instead, for any
positive integer $l$, stable fixed points exist from
$\gamma=1-\displaystyle \frac{K(a+\frac1{8a})}{2\pi l}=
1-\frac{0.77588......}l$ to $\gamma=1-\displaystyle
\frac{K(1-a)}{2\pi l}=1-\displaystyle \frac{0.51725......}l$
\cite{notelneg}.

\begin{figure}[ht]
\includegraphics[width=\columnwidth]{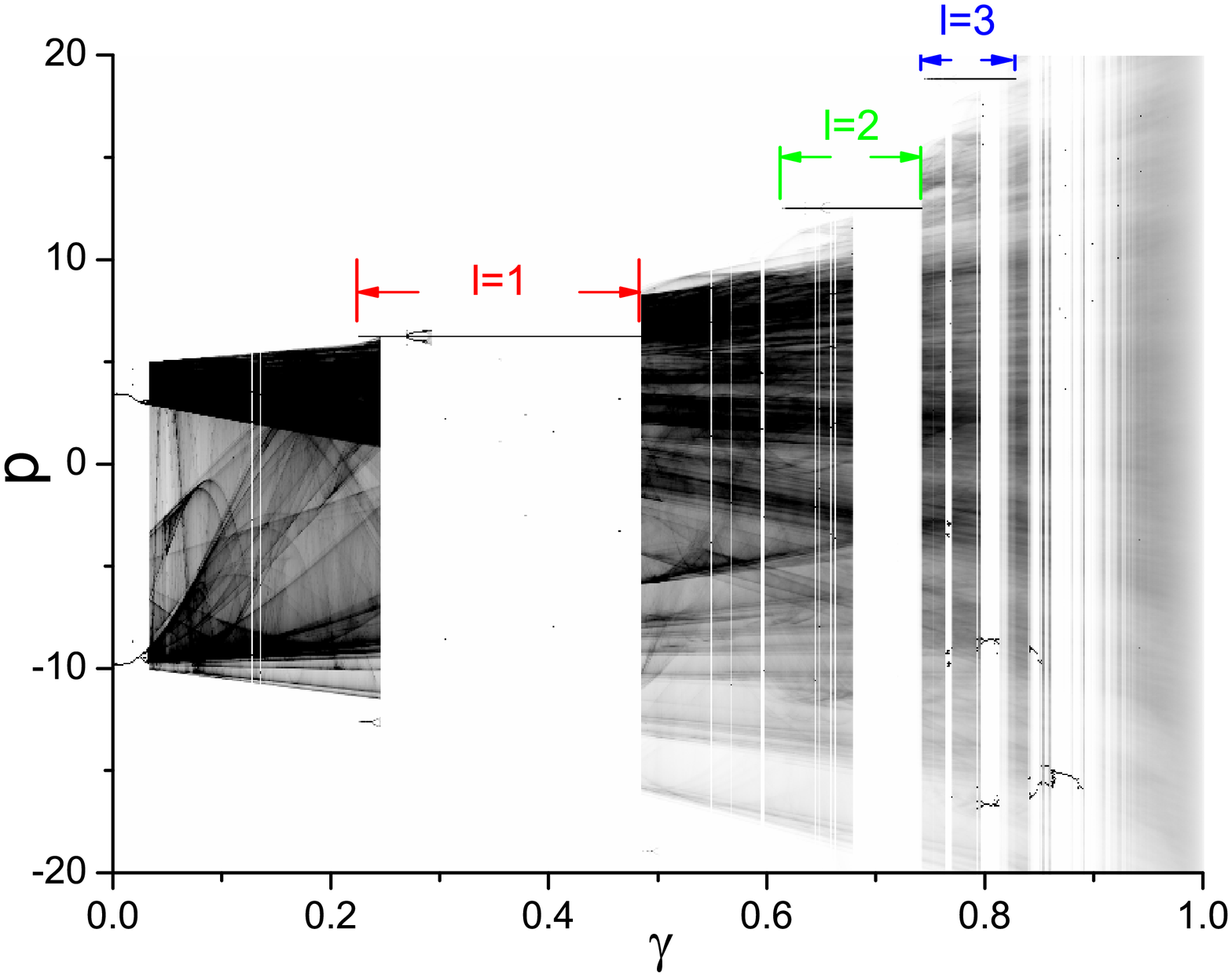}
\vspace{-1cm} \caption{\label{fig:bif1} Bifurcation diagram:
$5\times 10^3$ iterates are drawn after a transient of
$10^5$ map steps, starting
from 3600 initial states drawn from a uniform distribution
in the unit cell $0\le x<2\pi$, $-\pi\le p <\pi$.
The analytical expectations for the position of fixed point
windows at $a=0.5$ are also indicated.}
\end{figure}

The bifurcation diagram of Fig.~\ref{fig:bif1} confirms the above
analytical estimates. The positions of the the main fixed point
windows ($l=1,2,3$) coincide with the plateaus $\langle p \rangle =
2\pi,4\pi,6\pi$ observed in Fig.~\ref{fig:ensembleaveragep_k65a05}.
At both ends of the fixed point windows tangent bifurcations occur
which correspond to transitions from simple to strange attractors. (A
similar picture holds for stable periodic orbits, even though their
positions can not be so easily calculated analytically).

Since the width of the $l$-th stability window is proportional to
$\frac{1}{l}$ and $\sum_{l=1}^{\infty} \frac 1l$ does not converge,
these windows must overlap when $l$ is large enough. Therefore,
multiple attractors (and also periodic orbits) must coexist when
$\gamma$ approaches 1. Their attractive basins cut the phase space
into many pieces and here one may expect the asymptotic ratchet
current to depend, in general, on initial conditions. The large
negative values of $\langle p \rangle$ that appear at long times in
Fig.~\ref{fig:ensembleaveragep_k65a05} are the consequence of
trajectories ending up on periodic orbits after a long transient
chaotic motion.

\begin{figure}[ht]
\includegraphics[width=\columnwidth]{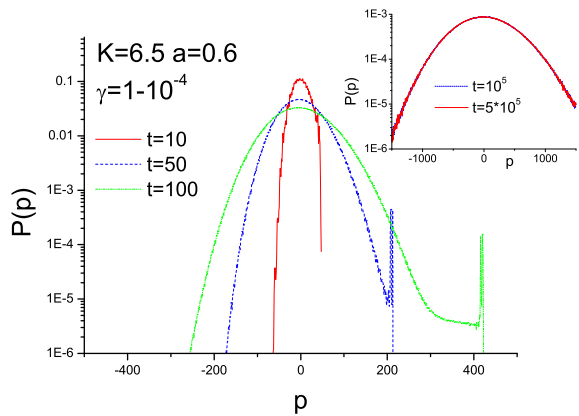}
\vspace{-1cm} \caption{\label{fig:pdisg0001} Snapshots of the
momentum probability distribution for $1-\gamma=10^{-4}$, starting
from $3.6\times 10^9$ initial conditions randomly and uniformly
chosen inside the unit cell $p\in[-\pi,\pi), x\in[0,2\pi)$. In the
inset the two curves drawn after long integration times, $t=10^5$
and $t=5\times 10^5$, overlap. This indicates that the distribution
saturates (at least up to the integration times explored in our
numerical simulations). }
\end{figure}

For very small dissipation ($\gamma\gsim 0.98$), we could only
observe chaotic motion on a fractal set even though it cannot be
excluded that this is a transient with a lifetime much exceeding the
total integration time. On the other hand this weakly dissipative
regime possesses very interesting features. Due to chaotic
diffusion, the momentum probability distribution
(Fig.~\ref{fig:pdisg0001}) widens in time and eventually, due to
dissipation, saturates to a stationary distribution close to a
Gaussian. This is clearly seen in Fig.~\ref{fig:pdisg0001} in which
the most remarkable feature is the small peak moving ballistically
in the direction of positive $p$. This peak is due to the presence
of small stability islands in the Hamiltonian limit $\gamma=1$ and
play a key role in the generation of the ratchet current.

\begin{figure}[ht]
\includegraphics[width=\columnwidth]{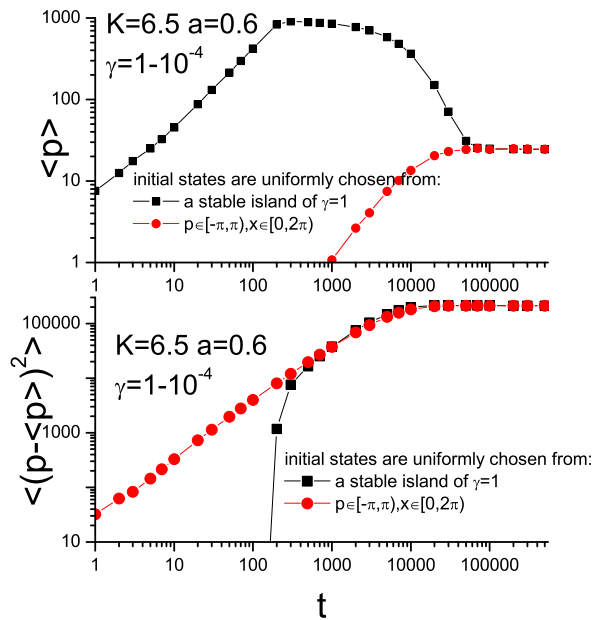}
\vspace{-1cm} \caption{\label{fig:avepdiffinitg0001}
First and second moment of the distribution $P(p)$ as a function of
time, for two different initial distributions.}
\end{figure}

To illustrate this point, we plot in
Fig.~\ref{fig:avepdiffinitg0001} the evolution in time of the first
two moments of the probability distribution $P(p)$, and compare two
different cases corresponding to two different initial conditions:
i) uniform distribution inside the unit cell $p\in[-\pi,\pi),
x\in[0,2\pi)$, ii) uniform distribution inside the main transporting
stability island. This island, for $\gamma=1$, corresponds to
ballistic motion in the positive momentum direction (``accelerator
mode''), with the momentum increased of $4\pi$ every $3$ map
iterations. Indeed this island is centered around the periodic orbit
$(x_1,p_1)$$\approx$$(1.6073,4.9802)\to (x_2,p_2)$$\approx$
$(2.9103,1.303)\to(x_3,p_3)$$\approx$$(2.9103,0)
\to(x_4,p_4)$$=$$(x_1,p_1$$+$$4\pi$$=$$p_1)$ (here $x$ and $p$ are taken modulo
$2\pi$). The area of the island $\approx 10^{-3}$, that is only
$3\times 10^{-5}$ of the available phase space. Notwithstanding the
accelerator mode, the ratchet current averages to zero in the
Hamiltonian limit: a sum rule exists \cite{ketzmerick} such that the
motion of the islands in the direction of positive momentum is
balanced by the motion of the chaotic sea in the opposite direction.
This sum rule no longer applies when dissipation is included since
the latter mixes the sets which are invariant in the Hamiltonian
limit, namely the islands and the chaotic sea. Still, if dissipation
is weak, these two sets remain essentially disconnected for a long
time scale $t_m \propto \frac{1}{1-\gamma}$. As a consequence, as
shown in Fig.~\ref{fig:avepdiffinitg0001}, for $t$$<$$t_m$ the ratchet
current $\langle p \rangle$$\approx$$0$ when starting from the entire unit
cell, while it grows linearly in time when the initial distribution
is concentrated inside the island. In this latter case, as expected,
the acceleration of the island is $\frac{2}{3}(2\pi)$, that is, the
island is shifted along $p$ by $2$ unit cells every $3$ map
iterations. At $t$$\sim$$t_m$, a momentum high enough is reached to
allow dissipation to drive particles outside the island. Then the
motion of these particles becomes chaotic and therefore the second
moment of the distribution suddenly increases, while the first
moment decreases due to dissipation. Correspondingly, for the
initial condition inside the entire unit cell, the ratchet current
starts to increase at $t$$\sim$$t_m$ until the asymptotic value is
reached. This provides direct numerical evidence that the ratchet
current is generated thanks to the presence of integrable islands in
the Hamiltonian limit, the island and the chaotic sea being finally
mixed by dissipation.

It is interesting to point out that while the ratchet current - i.e.
the first moment of the distribution $P(p)$ - is determined by a
sophisticated dynamical effect, the second moment can be understood
by means of a simple Fokker-Planck equation:
\begin{equation}
\frac{\partial P}{\partial t}=\frac{\partial}{\partial p}
\left(\frac{1}{2}D\frac{\partial P}{\partial p}\right)+
(1-\gamma)\frac{\partial}{\partial p}(pP),
\end{equation}
where $D\propto K^2$ is the diffusion coefficient. The stationary
solution to this equation is a Gaussian distribution of variance
$\frac{D}{2(1-\gamma)}$ which turns out to be in very good agreement
with our numerical data.

\begin{figure}[ht]
\includegraphics[width=\columnwidth]{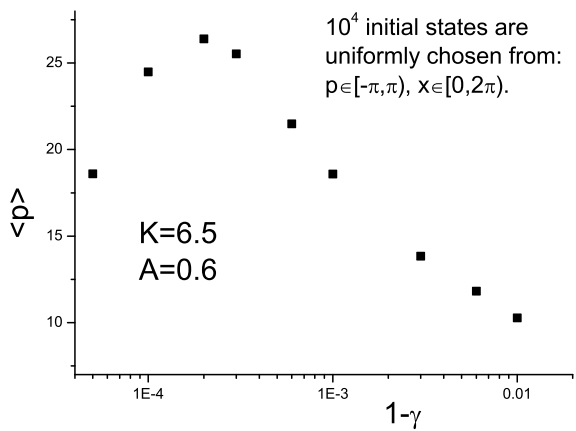}
\vspace{-1cm} \caption{\label{fig:pversuslargegamma} Stationary
ratchet current in the weak dissipation regime. The current is
averaged over $2\times 10^7$ map iterations, taken after an initial
transient of $2\times 10^7$ map steps.}
\end{figure}

We would like to stress that, thanks to the presence of integrable
islands in the Hamiltonian limit, large ratchet currents  can be
achieved also for weak dissipation: as shown in
Fig.~\ref{fig:pversuslargegamma}, we can have $\langle p
\rangle>26$.

A completely different behavior takes place when this structure
of islands is absent. Consider for example the ratchet
model:
\begin{align}
\left\{
 \begin{array}{l}
  \bar p=\gamma p+K[x-\pi-a\cos(x)], \\
  \bar x=x+\bar p.
 \end{array}
\right.
\end{align}
This map, in the Hamiltonian limit and for $a\in[0,1)$ is completely
chaotic with no stability islands \cite{liverani}.
For this map, we have always
found as stationary distribution a strange attractor, supporting a
very weak ratchet current. For instance, at $K=1$, $a=0.7$ we have
obtained $\langle p \rangle <0.13$ for any value of $\gamma$.
Basically the ratchet current is only due to the asymmetry of the
attracting set which, at least for weak dissipation, is also weak.

In summary, the results presented in this paper show that large
ratchet currents can be generated in a dissipative system thanks to
the presence, in the Hamiltonian limit, of transporting integrable
islands embedded in a chaotic sea. This phenomenon leads, due to the
joint presence of chaos and dissipation, to large ratchet currents nearly
independently of initial conditions and is generic for systems with
spatial and temporal periodicity, in that transporting islands are
typical for such systems in the Hamiltonian limit.

Finally, we point out that the dissipative ratchet model discussed
in this paper could be realized by means of cold atoms in optical
lattices, where it is possible to implement the
asymmetric potential (\ref{ratchetpotential}) \cite{weitz} and
a velocity proportional damping \cite{scully}.
Since kicked systems similar to our have been
implemented in the deep quantum regime \cite{raizen},
it appears possible to investigate experimentally
the impact of purely quantum effects such as dynamical localization
on the ratchet transport discussed in this paper.

We thank Gabriel Carlo for useful discussions. This work is
supported in part by an Academic Research Grant from MOE and the
DSTA under Project Agreement No. POD0410553.


\begin{thebibliography}{01}
\bibitem{Hlimit}
Note, however, that, in the Hamiltonian limit, the ratchet effect
in systems with mixed phase space dynamics is discussed in
\cite{ketzmerick,jangbin}.
\bibitem{notelneg}
Note that there also exist stable fixed points for negative
integers $l$ when $l\le -2$; however, we have observed in our
numerical simulations that their effect on the
ratchet current is very small.
\end{thebibliography}
\end{document}